# Anticipating or Speculating? Making Futures Pay in AI Ethics

John P. Nelson[1]

***Abstract*—Philosophers have recently been trying to develop uniform decision rules for instances in which physicians' judgments diverge from the recommendations of medical artificial intelligence (AI) systems. I argue that this effort is fatally flawed because the performance and appropriate use of algorithmic recommendation systems by physicians are highly context-specific, empirical medical matters rather than epistemological ones. In such instances, general decision rules can be stipulated only to the extent that they abstract away from relevant, indeed determinative, contextual circumstances. The alleged problem of physician-AI disagreement, and others like it, play into longstanding disputes in technology ethics about the viability and utility of envisioning possible futures—is this useful "anticipation" or wasteful "speculation"? I suggest that the two may be distinguished. Anticipation explores plausible futures to help inform present actions under irreducible uncertainty. Speculation abstracts away uncertainty, developing formal solutions to stipulated circumstances whose relationship to the real present is often quite unclear. Efforts to develop crisp philosophical solutions to physician-AI disagreement, among other problems, fall into the latter category. Technology ethics should remember that fruitful futures thinking must both begin from and return to the messy, uncertain present.**

[1]School of Public Policy, Oregon State University, Corvallis, USA

Email: nelsoj34@oregonstate.edu
ORCID: 0000-0002-3010-2046

> "Plans are nothing, but planning is everything. . . . [T]he first thing you do is to take all the plans off the top shelf and throw them out the window and start once more. But if you haven't been planning you can't start to work, intelligently at least. That is the reason it is so important to plan, to keep yourselves steeped in the character of the problem that you may one day be called upon to solve—or to help to solve."
>
> —Dwight D. Eisenhower (1957)

## 1. Anticipating What? Physician-AI Disagreement and "Speculative Ethics"

Recent years have seen substantial improvements in the efficacy of machine learning (ML)-based medical tools, particularly for diagnostics (Asif et al., 2024). Such tools can often rival or exceed the performance of human diagnosticians across areas such as visual inspection of potential tumor imaging, and strides are being made in diagnosis and prognosis for viral diseases, diabetes, internal organ diseases, and more. In keeping with the large language model-driven rebrand of ML, these tools are already referred to as artificial intelligence (AI). On the horizon are more sophisticated tools which might not only offer diagnoses, but recommend courses of treatment, for individual patients, based on inputs about the individual patient and the vast tranches of patient data on which the systems have been trained. AI systems can integrate possibly more, and certainly different, population-level data compared to doctors in generating such recommendations. This means that, at times, such systems could offer recommendations which diverge from those of the attending physician.

In these fascinating developments, bioethicists and epistemologists have smelled blood: the prospect of philosophical questions. ML-based systems can, in a certain way, draw inferences from an evidentiary base not accessible to any single practicing physician. Some philosophers have therefore characterized such systems as potentially independent "epistemic authorities" to physicians. This has allowed them to take "disagreements" between clinicians and AI systems as epistemological and moral-philosophical problems, which call for philosophical resolution (see Clough, 2003 for a critique of the tendency to "philosophize" practical problems). Philosophers have articulated a variety of principles which they hope might be used to resolve such quandaries. These have included that the physician should defer to the AI system (when it has been shown superior—we'll come back to this) (Bjerring & Busch, 2020; Grote & Berens, 2020); that physicians should (or "should be able to," which merely annuls the deference decision rule) reject the AI system's suggestion; or, quaintest of all, that such disagreement should require a third opinion from another human physician as the tiebreaker (Kempt & Nagel, 2022). It would be nice if a single decision rule could tell physicians what to do when they disagree with AI systems. But would it work out in practice?

Before engaging the substance of this question, it is worth situating it in the broader landscape of technology and innovation ethics. Many scholars, including myself, have advocated vociferously for efforts to foresee possible effects of emerging technologies on societal values, and to take action to align technological development and implementation with such values (Barben et al., 2008; Guston, 2014; Nelson & Bozeman, 2025; Nelson et al., 2022; Nelson et al., 2021; Ribeiro et al., 2018; Selin et al., 2023; Stilgoe et al., 2013). Such efforts have been critiqued, most prominently by Alfred Nordmann (Nordmann, 2007, 2014), as "speculative ethics": an incoherent engagement with hypostatized possible futures, which distracts from genuine and present problems. Related critiques have more recently been leveled against those versions of AI ethics which concern themselves with the possibilities of superintelligences and robot apocalypses rather than mundane and present problems of bias, inequity, and authoritarianism (Gebru & Torres, 2024; Munn, 2022; Prunkl & Whittlestone, 2020; Sætra & Danaher, 2023).

I follow Nordmann's opponents in arguing that appropriately humble efforts at anticipation, focused on the workaday problem of anticipating possible consequences and choosing robust actions under irreducible uncertainty, are not only essential but in many forms unavoidable (Guston, 2014; Selin, 2014; Selin et al., 2015). But it seems obvious that his critique could land in some cases. I think that the alleged problem of AI-physician disagreement is an example of an instance where it does. Rather than engaging with the uncertainties of medical AI development, implementation, and application, scholars writing on this problem have indeed hypostatized abstract, contextless situations with no clear relationships to the actual state of affairs. Doing this has, conveniently, allowed them to define problems which admit of neat philosophical resolution.

This strategy is far from unique to AI-physician relations. For example, I have previously critiqued it in the contexts of human genome editing and of synthetic life (Nelson et al., 2022; Nelson et al., 2021). Such cases may help to define some diagnostic criteria for "speculative ethics" in an appropriately pejorative sense. I suggest that speculative ethics tries to tell us exactly what we ought to do in hypothetical, but well-specified, situations which may or may not ever come to pass; whereas anticipatory governance properly tries to build our capacity to cope with real, present situations which cannot be completely known (Guston, 2014).

In the following section, I attempt to show the superfluity of the physician-AI disagreement problem by building upon a recent critique from Giorgia Pozzi and colleagues (Pozzi et al., 2026). In short, I argue that physicians have long consulted various sources of additional information, including reference texts, diagnostic tools, and even other physicians, without needing general decision rules; and, indeed, that the physician's expertise consists in nuanced understanding of how to contextually use such aids. In the final section, I expand this critique into a cautionary lesson for future-focused technology ethics in general. Anticipation should help us to cope with the messy present, rather than constructing imagined, simplified futures for neat philosophical disposal.

## II. No Algorithmic Agents: Physician-AI Disagreement as Pseudoproblem

In a recent article in the *American Journal of Bioethics*, Giorgia Pozzi and colleagues show that none of the three decision rules proffered for physician-AI disagreement—deference, overruling, and second opinion—can be made to work in practice (Pozzi et al., 2026). Against uniform deference, they observe that AI system-based decisions cannot necessarily be made epistemically accountable to patients; that the superiority of their performance may be overstated and difficult to ascertain; and in particular that an ML system may be incapable of integrating specific information about patient history, symptoms, preferences, and risk tolerance. Against uniform overruling, they observe that a physician might usefully modify her treatment plans in response to an AI system rather than overruling it outright, and that universal overruling might undercut the utility of such ML systems in general. And against uniform requirements for a second opinion, they observe that such a requirement might not be possible to fulfill in time-sensitive care and might add yet further layers of disagreement rather than resolving it. All these critiques seem at least sometimes compelling—though perhaps not always. I suggest that the problem with the critiqued positions is less their substance than the expectation of uniformity. Indeed, Pozzi and colleagues suggest that we should stop regarding "disagreement" between AI systems and doctors as a problem, and should instead make space for various nuanced and contextual methods of resolution (Pozzi et al., 2026).

I would push this argument somewhat further. I don't see the point of describing divergence between AI system recommendations and physicians as "disagreements" at all. This section's title is meant in two senses. First, I think that regarding recommendation-generating AI systems as even candidates for epistemic authority, or other such agential categories, produces confusion rather than benefit. AI system recommendations should be treated like blood test results: evidence to be evaluated by a responsible physician. It is for the physician to learn and judge how to use this new and elaborate variety of evidence alongside the older ones of sensory perception, medical literature, and diagnostic instruments (Schon, 1984). Second, any decision rule which can be made generally defensible will abstract away from the details of particular patients, AI systems, and the interrelations between the two. But, as Pozzi and colleagues show, it is exactly these fine details that determine how one should judge an algorithmically generated recommendation in practice. In short, there is no reason to use agential language about algorithmic recommendation systems unless and until we accord them some of the rights and responsibilities of independent medical agents. And the hope of developing a priori, algorithmic guidance for physicians about how to use algorithmic recommendation systems is a vain one.

Here's a way to cultivate this intuition. Imagine a medical malpractice suit that turned on a physician's choice to disregard the result of a blood test. She made this judgment based on her reading of the contextual details of the situation: patient symptoms and statements, her prior experience with and reading about the test, and more. What standards would be used to decide this suit? It is likely that the physician's choice would be compared to "best practice" in her field, as articulated by major medical associations and by physicians called as expert witnesses. "Lay expert" patient advocates with expertise germane to the test might be called as well, as might lab technicians and engineers for the manufacturer. This makes sense, as all these various actors have knowledge of the particular test, its capabilities and limitations, and how the medical community judges it to be best used in practice.

Suppose now that a moral philosopher had previously written a paper articulating a decision rule for when physicians ought to accept the results of blood tests in general. Aiming for an uncontestable principle, she stated that physicians ought to accept blood test results when the test has been shown to be accurate for the relevant application by a long history of trials, and has been conducted competently. But, of course, the questions on which the malpractice suit will hang are precisely those which the philosopher has assumed away. Was the test indeed relevant and accurate for the application to which it was put? Was it competently conducted? Did the defendant behave appropriately in her judgments about these things? These questions are not only relevant to the suit; they are the suit. Would this moral philosopher be called to testify? Should she be? Does she have additional insight or information germane to the case? It is no slight against philosophers to suggest that the answer to all these questions is "no." Subject-matter expertise on narrow diagnostic and treatment questions is the domain of physicians, not ethicists.

I suggest we should view AI diagnostic or treatment recommendation systems along the model of the blood test in this scenario, rather than as potentially independent epistemic authorities about which general decision rules might be formulated. No one would ever dream of articulating a general decision rule about when to use blood test results to determine diagnosis or treatment. They would recognize that the appropriate decision necessarily depends upon the contextual relevance and accuracy of the blood test. Any generally acceptable rule on this topic would be so abstract as to give no guidance in particular cases.

Though AI recommendation systems are far more complicated than blood tests, I do not see why we should treat them differently from other evidence sources. They should be regarded as specific tools and evaluated individually, rather than promoted to independent epistemic entities requiring treatment as a general class. The questions of when, where, and how to employ any particular diagnostic technique or treatment planning instrument, and when and how each is reliable or unreliable, are valid and important. But these are practical, specific, empirical questions. The core work of physicians is developing the best answers they can to questions like these (Allen, 2020; Porter, 1997). Due to limitations and biases in training data and methods, it is almost inconceivable that an AI system's relevance and accuracy should be the same across all patient subpopulations and situations (Huang et al., 2022; Otokiti et al., 2025). To apply a particular AI system effectively, the physician will have to use her own experience and knowledge of its contextual relevance, scope, and accuracy. It is of little use to generate decision rules based on stipulated degrees and domains of AI system accuracy, for the hard part is judging when and where such degrees and domains obtain.

Concepts of best practice are entirely sufficient to require the use of some particular tool, if the medical community judges it reliable enough. Why should AI recommendation systems require more? No question of responsibility has been raised; it is almost never suggested that AI systems should be given responsibility for final patient care decision-making rather than physicians or patients. Nor has any question of authority been raised; no one has suggested that AI systems have rights to be used or to govern patients or physicians. In the absence of such suggestions, it is not clear what it even means to suggest that AI systems may be independent epistemic authorities rather than evidence sources like all the rest.

AI systems are novel, but not in any way that need alter the basic structure of physician autonomy and responsibility. AI systems can provide contextual recommendations distilled from vast evidentiary bases beyond the physician's personal knowledge and experience. But so can reference texts, medical literature, and flowcharts. No one has suggested that these tools, or indeed even the genuine and obvious independent epistemic authorities of consulting colleagues, require general philosophical decision rules for use. As of summer 2026, the American Medical Association's (2026) Code of Medical Ethics contains no formal guidance as to exactly how much weight an attending physician should place on a consulting physician's recommendations. Nor does there appear to be a thriving bioethical literature on this subject. The problem of physician disagreement with a colleague is the very model of the problem now raised with respect to AI systems. Yet a requirement that the attending physician should always or never defer to a consulting physician, or should always seek yet another opinion, is rightly recognized as absurd. No one would presume that the question of how much weight to put on a consulting colleague's recommendation can be answered in general terms. Nothing is gained by asking this question about AI systems, either.

The question which Pozzi and colleagues so effectively vitiate rests upon two key errors. The first is to assume that a novel information technology, because complicated and dynamic, must merit qualitatively distinct treatment from older ones. The second is to assume that instances of this technology constitute a coherent category which can be usefully treated in general and philosophical terms, rather than, like all other medical tools, in specific and empirical ones. To seek a general solution for how to deal with this new variety of evidence across all cases and contexts is to ask too much of philosophy. I do not think it would even occur to us to do this if not for our tendency to anthropomorphize complicated computational systems.

To put it playfully, we ought to institute a rule of deference of ethicists to doctors as pertains to diagnostic and treatment standards--topics on which physicians are definitionally and manifestly superior epistemic authorities to ethicists. It is for the expert medical community to evaluate the particular utility and appropriateness of particular AI systems, both across and within specific cases. In more familiar domains such as blood test interpretation and colleague consultation, such questions are rightly recognized as hanging on the fine details of the situation, and therefore as being the physician's job. In our dappled and dynamic world of heterogeneity, uncertainty, and trade-offs, there can be no general algorithm for what to do with algorithmically produced recommendations.

## III. Lessons for Responsible Innovation: Avoiding Speculative Ethics

I hope to have convinced the reader that the alleged problem of physician-AI disagreement need be neither raised nor treated in philosophical terms. There is a great deal of important work to be done in aligning development and implementation of medical AI with societal needs and values, including extensive empirical study of the characteristics of particular medical AI systems. But, shorn of any concrete context, the question of what to do when AI systems' outputs diverge from physicians' judgments becomes a mere word game, a situation so artificial as to obscure the absurdity of seeking a single solution across this vast and heterogeneous landscape of interactions.

This is far from the only instance in which such word games occur. AI ethics is rife with discussion of the potential moral status of computer systems, but discussion of the operational implications of such moral status is comparatively scant (Nelson, 2026). "Longtermist" thinking attempts to factor future persons and technologies into morality and decision-making, but brushes over manifold assumptions and uncertainties in the factoring (Wenar, 2024).

Indeed, focus on tractable thought experiments to the exclusion of reckoning with uncertainty is a much longer-standing phenomenon in technology ethics generally. Years ago, I critiqued the German Ethics Council's (2019) mammoth consensus report on human germline genome editing for analyzing only "an anonymous, typified subject of intervention; the intervention method itself; and an unexplored, presumed social backdrop . . . [severed] from the particularities of possible contexts." The Council's approach, I argued, "exclud[ed] from exploration many or most of the factors which might determine the roles [human germline genome editing] would play with respect to . . . distributional or procedural justice, the freedom of persons [edited] and persons not, and social solidarity or marginalization," despite characterizing these outcomes as very important (Nelson et al., 2021, p. 398). Such exclusions were the price of being able to definitively conclude the in-principle permissibility of various human germline genome editing applications.[1] But applications never occur only in principle. Similarly, I critiqued Ankeny and colleagues' (2022) generally praiseworthy advocacy for anticipatory methods around synthetic life for a narrow focus on the in-principle ethical status of entities which might or might not be created in the lab, rather than on the behavior and goals of the research, development, and commercial institutions which determine such entities' societal consequences (Nelson et al., 2022). In such cases we can see a mode of futures thinking focused on the generation of a neat problem that admits of a definitive solution, even at the cost of realism or relevance.

Yet the problem of aligning technological development and implementation with societal needs and values is inherently a messy and contested one (Ribeiro et al., 2023)—"wicked," in today's overused terminology (Alford & Head, 2017). In-principle solutions to simplified, fully characterized problems can avail us little when the basic contours of our situations are uncertain and contested (Allenby & Sarewitz, 2011; Lindblom, 1959, 1979, 1986). It may well be that physicians should defer to superior AI systems. But when, where, and how are AI systems superior? Perhaps human germline genome editing is sometimes ethically permissible in principle. But what will its consequences be in a world of inequality, injustice, and institutions in general? Perhaps we would be obligated to prioritize computer science and space exploration above ameliorating the suffering of people actually living, if we could be sure our investment would enable trillions of unborn future persons to thrive. But we cannot achieve such surety (Wenar, 2010, 2024). The kind of debate-club ethics exemplified in the first half of each of these dyads may be properly dismissed as "speculative." It does not try to help us determine how, provisionally, contingently, uncertainly, to protect and advance our values amid a blooming, buzzing profusion of possible futures. Instead, it imagines situations which admit of clear answers, achieved by begging any number of empirical questions.

Anticipation, done well, aims not to abstract away uncertainty, but to help us cope with it. Scenario planning explicitly begins with what is contingently known about the present state of affairs, projects it forward into several different plausible futures, and attempts to identify actions and strategies which will be robust to variation across those scenarios (Schwartz & Ogilvy, 2004; Selin et al., 2023; Withycombe Keeler & Bernstein, 2021; Withycombe Keeler et al., 2019). Backcasting defines desired states of the future and attempts to work backwards through buffeting uncertainty to what can be done today (Withycombe Keeler et al., 2022). Threatcasting envisions possibly dangerous future situations and identifies signs that they may be coming about (Johnson et al., 2021). All these methods use possible futures to try to define robust actions and approaches which can be taken under today's situation of limited knowledge and contested values, rather than to produce in-principle solutions to crisp problems which may never come about.

Over a century ago, John Dewey could take for granted that "ready-made rules available at a moment's notice for settling any kind of . . . difficulty," were in medicine "known as quackery" (Dewey, 1922, p. 238). He argued that the same attitude should be imported to philosophy. Unfortunately, it seems that the reverse has occurred. Philosophy has essential roles to play in summarizing our aspirations and practices, seeking ways to reconcile them, and clarifying our disagreements (Dewey, 1920; Rorty, 2021). Anticipation, however, should produce not syllogisms but strategies. Strategies are provisional, openly contingent, value-laden, of uncertain reliability--and relevant to decision-makers embedded in some concrete, messy situation with horizons of values, capacities, and knowledge. Thinking about the future should help us to better understand and cope with the present before us. Anything less than this is speculation indeed.

There is a caveat, of course. Speculation can help us to expand our horizons, envision worlds unseen, and confront again forgotten elements of our environments, communities, selves, and deeds (Le Guin, 1979; Rorty, 1989; Shippey, 2000; Tolkien, 1939 [2014]). But it does so most effectively when embedded in concrete situations, which force the author to confront the messiness and uncertainty of human lives and institutions (Finn, 2025). Speculation can be of great value—so long as it can be kept honestly fictional. Paradoxically, this may help it both to start from and to return to the present and the real.

[1] This approach is in line with the general reductionist strategy of Western philosophy since Plato: start with a simple problem that admits of an uncontested answer, infer general principles, and attempt to reconstruct a science of morality. Such exercises always leave an arbitrary amount of room for divergence in detailed application--when applications are included at all. In the real world, simple situations are not elementary particles of a moral physics to come, but contingent products of rhetoric, long habit, or elaborate institutional scaffolding. Judgment and discretion can be deferred, submerged, or redistributed, but not eliminated (Merry 2016; Rorty 2021; Scott 1998; Stone 2011; Walzer 1994).